\documentstyle[preprint,aps,prb,epsf]{revtex}
\begin{document}
\preprint
\widetext
\title{Strong-coupling expansions for the pure and disordered bose
Hubbard model}
\author{
J. K. Freericks$^{(a)}$ and H. Monien$^{(b)}$.
}
\address{
$^{(a)}$Department of Physics,
Georgetown University, Washington, DC 20057\\
$^{(b)}$Theoretische Physik, ETH H\"onggerberg,
CH-8093 Z\"urich, Switzerland
}
\date{\today}
\maketitle
\widetext
\begin{abstract}
A strong-coupling expansion for the phase boundary of the (incompressible)
Mott insulator is presented for the bose Hubbard model.  Both the pure case
and the disordered case are examined.  Extrapolations of the series expansions
provide results that are as accurate as the Monte Carlo simulations and agree
with the exact
solutions.  The shape difference between Kosterlitz-Thouless critical behavior
in one-dimension and power-law singularities in higher dimensions arises
naturally in this strong-coupling expansion. Bounded disorder distributions
produce
a ``first-order'' kink to the Mott phase boundary in the thermodynamic limit
because of the presence of Lifshitz's rare regions.
\end{abstract}
\narrowtext
\section{Introduction}
Strongly interacting bosonic systems have attracted a lot of recent
interest\cite{Fishers,Scalettar,Rokhsar,Girvin}.
Physical realizations
include short correlation-length superconductors, granular
superconductors, Josephson arrays, the dynamics of flux lattices
in type II superconductors, and critical behavior of $^4{\rm He}$ in
porous media.
The bosonic systems are either tightly bound composites
of fermions that act like
effective bosonic particles with soft cores, or correspond to
bosonic excitations that have repulsive interactions.  For this reason, these
systems are modeled by soft-core bosons which are described most simply
by the bose Hubbard model.
Various aspects of this model were investigated analytically by
mean-field theory \cite{Fishers,KampfZimanyi}, by renormalization group
techniques \cite{Fishers,Rokhsar} and by projection methods \cite{Krauth0}.
The bose Hubbard model has also been studied with quantum Monte
Carlo methods (QMC) by Batrouni et al.  \cite{Scalettar}
in one dimension (1+1)
and by Krauth and Trivedi \cite{Krauth3}, van Otterlo and
Wagenblast\cite{vanOtterlo}, and Batrouni et al.\cite{Batrouni}
in two dimensions (2+1).
In this contribution, the Mott phase diagram is
obtained from a strong-coupling expansion that
has the correct dependence on spatial dimensionality, is as accurate as the
QMC calculations, and agrees with the known exact solutions. Preliminary
results for the pure case have already appeared\cite{Freericks_monien}.

The bose Hubbard model is the
minimal model which contains the key physics of the strongly
interacting bose systems---the competition between kinetic and
potential energy effects.  Its Hamiltonian is
\begin{equation}
  H = - \sum_{ij} t_{ij} b^\dagger_i b^{\phantom{\dagger}}_j+\sum_i\epsilon_i
{\hat n}_i
   - \mu \sum_i {\hat n_i} + \frac{1}{2} U \sum_i {\hat n}_i ({\hat n}_i-1)
  \quad , \quad
  {\hat n}_i =  b^\dagger_i b^{\phantom{\dagger}}_i
  \label{H}
\end{equation}
where $b_i$ is the boson
annihilation operator at site $i$, $t_{ij}$ is the hopping matrix element
between the site $i$ and $j$, $\epsilon_i$ is the local site energy,
$U$ is the strength of the on-site repulsion, and $\mu$ is the chemical
potential. The hopping matrix is assumed to be
a real symmetric matrix $(t_{ij}=t_{ji})$ and the lattice is also assumed to
be bipartite; {\it i.~e.},
the lattice may be separated into two sublattices (the $A$ sublattice and the
$B$ sublattice) such that $t_{ij}$ vanishes whenever $i$ and $j$ both
belong to the same sublattice (in particular, this implies $t_{ii}=0$).
The local site energy $\epsilon_i$ is a quenched random variable
chosen from a distribution of site energies that is symmetric about zero and
satisfies $\sum_i\epsilon_i=0$.  The pure case corresponds to all site energies
vanishing $(\epsilon_i=0)$.

The form of the zero temperature ($T=0$) phase diagram can be understood
by starting from the strong-coupling or ``atomic'' limit
\cite{Fishers,GiamarchiSchulz,Ma}.
In this limit, the kinetic energy vanishes ($t_{ij} = 0$)
and every site is occupied by a fixed
number of bosons, $n_0$. In the pure case, the ground-state boson
occupancy ($n_0$) is the same for each lattice site, and
is chosen to minimize the on-site energy.
If the chemical potential, $\mu=(n_0+\delta)U$, is parameterized
in terms of the deviation, $\delta$, from
integer filling $n_0$, then the on-site energy is
$E(n_0) = -\delta U n_0 - \frac{1}{2} U n_0 (n_0 + 1)$, and
the energy to add a boson
onto a particular site satisfies $E(n_0+1) - E(n_0) = -\delta U$.
Thus for a nonzero $\delta$, a finite amount of energy is required to
move a particle through the lattice.
The bosons are incompressible and localized, which produces
a Mott insulator. For $\delta = 0$, the ground-state
energies of the two different boson densities are degenerate
[$E(n_0) = E(n_0+1)$] and no energy is needed
to add or extract a particle; {\it i. e.}, the compressibility is finite and
the system is a conductor.
As the strength of the hopping matrix elements increases, the range of
the chemical potential $\mu$ about which the system is incompressible
decreases.
The Mott-insulator phase will completely disappear at a critical value of the
hopping matrix elements. Beyond this critical value of $t_{ij}$ the system
is a superfluid.

In the disordered case, a Mott-insulating phase may or may not exist depending
upon the strength of the disorder.  The energy to add a boson onto site $i$
becomes $E(n_0+1)-E(n_0)=\epsilon_i-\delta U$, so that the system is
compressible if a site $i$ can be found which satisfies $\epsilon_i=\delta U$.
If the disorder is assumed to be
symmetrically bounded about zero ($|\epsilon_i|\le \Delta U$) then a Mott
insulator exists whenever $\Delta <{1\over 2}$.  The ground-state boson
occupancy is uniformly equal to $n_0$ within the Mott insulating phase
which extends from $-\Delta\ge \delta\ge\Delta-1$ (when $t_{ij}=0$).
Once again, the bosons are incompressible within the Mott
phase and the system is insulating.  As the hopping matrix elements increase
in magnitude, the range of the chemical potential within which the system
is incompressible decreases until the Mott phase vanishes at a critical value
of the hopping matrix elements.
The compressible phase will typically also
be an insulator and is called a bose glass\cite{Fishers}, but it has been
conjectured that
in some cases the transition proceeds directly from the Mott insulator
to the superfluid\cite{Fishers,Rokhsar}.

The phase boundary between the incompressible phase (Mott insulator)
and the compressible phase (superfluid or bose glass) is determined here in
a strong-coupling expansion by calculating both
the energy of the Mott insulator and of a defect state (which contains
an extra hole or particle) in a perturbative expansion of the single-particle
terms $-\sum_{ij}t_{ij}b_i^{\dag}b_j+\sum_i\epsilon_i{\hat n}_i$.
At the point where the energy of the Mott state is degenerate with the defect
state, the system becomes compressible.  In the pure case, the compressible
phase is also superfluid, but in the disordered case, the compressible
phase is a bose glass (except possibly at the tip of the Mott
lobe)\cite{Fishers,Rokhsar}.

There are two distinct cases for the defect state: $\delta < 0$ corresponds to
adding a {\it particle} to the Mott-insulator phase (with
$n_0$ bosons per site); and $\delta>0$ corresponds to adding a {\it hole} to
the
Mott-insulator phase (with $n_0+1$) bosons per site. Of course,
the phase boundary depends upon the number of bosons per site, $n_0$, of the
Mott insulator phase.

To zeroth order in $t_{ij}/U$ the Mott-insulating state is given by
\begin{equation}
  |\Psi_{\text{Mott}}(n_0)\rangle^{(0)} =
  \prod_{i=1}^N
  \frac{1}{\sqrt{n_0!}}\left(b^\dagger_i\right)^{n_0}|0\rangle
\end{equation}
where $n_0$ is the number of bosons on each site, $N$ is the number of
sites in the lattice and $|0\rangle$ is the vacuum state.
The defect state is characterized by one additional particle (hole) which moves
coherently throughout the lattice. To zeroth order in the single-particle terms
the wave function for the ``defect'' state is determined by degenerate
perturbation theory:
\begin{eqnarray}
  |\Psi_{\text{Def}}(n_0)\rangle^{(0)}_{\text{part}} &=&
    \frac{1}{\sqrt{n_0+1}} \sum_i f_i^{\rm (part)} b^\dagger_i
    |\Psi_{\text{Mott}}(n_0)\rangle^{(0)}
    \cr
  |\Psi_{\text{Def}}(n_0)\rangle^{(0)}_{\text{hole}} &=&
    \frac{1}{\sqrt{n_0}} \sum_i f_i^{\rm (hole)} b^{\phantom{\dagger}}_i
    |\Psi_{\text{Mott}}(n_0)\rangle^{(0)}
\end{eqnarray}
where the $f_i$ is the eigenvector of the corresponding single-particle
matrix $S_{ij}^{\rm (part)}(n_0)\equiv
-t_{ij}+\delta_{ij}\epsilon_i/(n_0+1)$
[$S_{ij}^{\rm (hole)}(n_0)\equiv -t_{ij}-\delta_{ij}\epsilon_i/n_0$] with the
lowest eigenvalue (the hopping matrix is assumed to have a nondegenerate
lowest eigenvalue).  It is well known that the minimal eigenvalue of the
single-particle matrix $S_{ij}$ is larger than the sum of the minimal
eigenvalue of the hopping matrix plus the minimal eigenvalue of the disorder
matrix.  However, it has been demonstrated that as the system size becomes
larger and larger, the minimal eigenvalue approaches the sum of the minimal
eigenvalues of the hopping matrix and of the disorder matrix as closely as
desired\cite{Lifshitz} (because of the existence of arbitrarily large
``rare regions'' where the system looks pure with $\epsilon_i=- \Delta U$
or with $\epsilon_i=\Delta U$).
Therefore, in the thermodynamic limit, the perturbative
energy of each defect state becomes
\begin{equation}
E_{\rm Def}^{\rm (part)}(n_0)-E_{\rm Mott}(n_0)=-\delta^{\rm (part)}U+
\lambda_{\rm min}(n_0+1)-\Delta U+...
\label{eq: particle first order}
\end{equation}
\begin{equation}
E_{\rm Def}^{\rm (hole)}(n_0)-E_{\rm Mott}(n_0)=\delta^{\rm (hole)}U+
\lambda_{\rm min}n_0-\Delta U+...
\label{eq: hole first order}
\end{equation}
to first order in $S$,
where $\lambda_{\rm min}$ is the minimal eigenvalue of the hopping matrix
$-t_{ij}$.  In the case of nearest-neighbor hopping
on a hypercubic lattice in $d$-dimensions, the number of nearest
neighbors satisfies $z=2d$ and the minimal eigenvalue is $\lambda_{\rm min}=
-zt$.

The boundary between the incompressible phase and the compressible
phase is determined when the energy difference between the Mott insulator
and the defect state vanishes (the compressibility is assumed to
approach zero continuously at the phase boundary).  Thus two branches
of the Mott lobe can be found depending upon whether the defect state is
an additional hole or an additional particle.
The two branches of the Mott-phase boundary meet when
\begin{equation}
  \delta^{\rm (part)}(n_0) + 1 = \delta^{\rm (hole)}(n_0).
  \label{eq: critical condition}
\end{equation}
The additional one on the left hand side
arises because $\delta$ is measured from the point
$\mu/U = n_0$.
Equation (\ref{eq: critical condition})
may be used to estimate the critical value of
the hopping matrix element beyond which no Mott-insulator phase exists.  Let
$x$ denote the combination $dt/U$ and consider the first-order expansions in
Eqs. (\ref{eq: particle first order}) and (\ref{eq: hole first order}).
The critical value of $x$ satisfies
\begin{equation}
x_{\rm crit}(n_0)={1-2\Delta\over 2(2n_0+1)}\quad ,
\label{eq: crit cond first order}
\end{equation}
which vanishes when the disorder strength becomes too large ($\Delta\ge 1/2$).
Note that the critical value of $x$ is {\it independent} of the dimension
of the lattice; {\it the dimensionality first enters at second
order in $t$.}  The slope of the phase boundaries about the point
$\mu=n_0 U$ are equal in magnitude
$[\lim_{x\rightarrow 0}\frac{d}{dx}\delta^{\rm (part)}(n_0,x) =
- \lim_{x\rightarrow 0}\frac{d}{dx}\delta^{\rm (hole)}(n_0+1,x) ]$,
but change their magnitude as a function of the density $n_0$,
{\it implying that the Mott-phase lobes always have an asymmetrical shape}.
Note
further that the presence of disorder shifts the phase boundaries uniformly
by $\Delta$, but the slope is {\it independent} of the disorder distribution.

The bose Hubbard model in the absence of disorder
is examined by a strong-coupling expansion through
third order in the single-particle matrix $S$ in Section II.  The exact
solution
for an infinite-dimensional lattice\cite{Fishers} is examined and various
different extrapolation techniques are employed that do and do not utilize
additional information of the scaling analysis of the critical point.  Section
III describes the similar results for the disordered bose Hubbard model and
a discussion follows in Section IV.

\section{The pure case}

The bose Hubbard model in Eq. (\ref{H}) is studied in the absence of disorder
($\epsilon_i=0$).  The many-body version of Rayleigh-Schr\" odinger
perturbation
theory is employed throughout.
To third order in $t_{ij}/U$, the energy of the Mott state with
$n_0$ bosons per site becomes
\begin{equation}
  E_{\text{Mott}}(n_0) = N
  \left[
  -\delta U n_0 - \frac{1}{2}U n_0 (n_0+1) -
\frac{1}{N}\sum_{ij}\frac{t_{ij}^2}
{U}n_0(n_0+1)
  \right]
  \label{eq: EMott}
\end{equation}
which is proportional to the number of lattice sites $N$.
Note that the odd-order terms in $t_{ij}/U$ vanish in the above
expansion (odd-order terms may enter for nonbipartite
lattices).  The energy difference between the Mott insulator
and the defect state with an additional particle ($\delta < 0$) satisfies
\begin{eqnarray}
  E_{\text{Def}}^{\rm (part)}(n_0) &- E_{\text{Mott}}(n_0) =
  -\delta^{(\text{part})} U+\lambda_{\rm min}(n_0+1)
  +{1\over 2U}\sum_{ij}t_{ij}^2f_j^2n_0(5n_0+4)
  -{1\over U}\lambda_{\rm min}^2n_0(n_0+1)\cr
  &+{1\over U^2} n_0(n_0+1)
  \Biggr[(2n_0+1)\lambda_{\rm min}^3-({25\over 4}n_0+{7\over 2})
  \lambda_{\rm min}\sum_{ij}t_{ij}^2f_j^2-(4n_0+2)\sum_{ij}
  f_it_{ij}^3f_j\Biggr]
  \label{eq: Edef upper}
\end{eqnarray}
to third order in $t_{ij}/U$; while the energy difference
between the Mott insulating phase and
the defect phase with an additional hole ($\delta > 0$) satisfies
\begin{eqnarray}
  E_{\text{Def}}^{\rm (hole)}(n_0) &- E_{\text{Mott}}(n_0)=
  \delta^{(\text{hole})}U+\lambda_{\rm min}n_0
  + {1\over 2U}\sum_{ij}t_{ij}^2f_j^2(n_0+1)(5n_0+1)
  -{1\over U}\lambda_{\rm min}^2n_0(n_0+1)\cr
  &+{1\over U^2} n_0(n_0+1)
  \Biggr[(2n_0+1)\lambda_{\rm min}^3-({25\over 4}n_0+\frac{11}{4})\lambda_{\rm
  min}\sum_{ij}t_{ij}^2f_j^2-(4n_0+2)\sum_{ij}f_it_{ij}^3f_j\Biggr]
  \label{eq: Edef lower}
\end{eqnarray}
The eigenvector $f_i$ is the minimal eigenvector of the hopping matrix
$-t_{ij}$
with eigenvalue $\lambda_{\rm min}$ and is identical in the particle and
hole sectors.  These results have been verified by small-cluster calculations
on two and four-site clusters.
Note that the energy difference in Eqs. (\ref{eq: Edef upper}) and
(\ref{eq: Edef lower}) is {\em independent} of the lattice size $N$ indicating
that QMC simulations should not have a very strong dependence on the lattice
size.

In the case of nearest-neighbor hopping on a
$d$-dimensional hypercubic lattice, the minimum eigenvalue satisfies
$\lambda_{\rm min}=-zt$, the sum $\sum_{ij}t_{ij}^2f_j^2$ becomes $zt^2$, and
the sum $\sum_{ij}f_it_{ij}^3f_j$ equals $zt^3$.
Equations (\ref{eq: Edef upper}) and (\ref{eq: Edef lower})
can then be solved for the shift in the chemical potential $\delta$ at which
the
system becomes compressible as a function of the parameter $x\equiv dt/U$.  The
results for the upper boundary are given by
\begin{eqnarray}
  \delta^{(part)}(n_0,x)
  &= - 2x ( n_0 + 1)  +{1\over d}x^2n_0(5n_0+4)-4x^2n_0(n_0+1)\cr
&+2x^3n_0(n_0+1)
\left [ (-8+{25\over 2d}-\frac{4}{d^2})n_0+(-4+{7\over d}-{2\over d^2})
\right ]
\label{eq: upper boundary}
\end{eqnarray}
to third order in $x$,
and the lower boundary is given by
\begin{eqnarray}
  \delta^{(hole)}(n_0, x)
  &= 2xn_0-{1\over d}x^2(n_0+1)(5n_0+1)+4x^2n_0(n_0+1)\cr
&-2x^3n_0(n_0+1)
\left [ (-8+{25\over 2d}-{4\over d^2})n_0+(-4+{11\over 2d}-{2\over d^2})
\right ],
  \label{eq: lower boundary}
\end{eqnarray}
to third order in $x$.

As a further check on the accuracy of the Mott phase boundaries in Eqs.
(\ref{eq: upper boundary}) and (\ref{eq: lower boundary}), we compare the
perturbative expansion to the exact solution on an infinite-dimensional
hypercubic lattice\cite{Fishers} (which corresponds to the mean-field
solution).  Note that the solution in Ref.~\onlinecite{Fishers} was for the
infinite-range-hopping model; this solution is {\it identical} to that on an
infinite-dimensional lattice in the pure case.
The Mott phase boundary may be expressed as
\begin{equation}
{\mu\over U}-n_0=-\frac{1}{2}-x\pm \sqrt{x^2-x(2n_0+1)+\frac{1}{4}}
\label{eq: inf-d delta}
\end{equation}
where the plus sign denotes the upper branch to the Mott lobe
$(\delta^{\rm (part)})$, and the
minus sign corresponds to the lower branch $(\delta^{\rm (hole)}-1)$.  The
critical point can also
be determined as the value of $x$ where the square root vanishes.
One finds
\begin{equation}
x_{\rm crit}=n_0+\frac{1}{2}-\sqrt{n_0(n_0+1)}
\label{eq: inf-d xcrit}
\end{equation}
which depends on $n_0$ as $1/n_0$ in the limit of large $n_0$.
The strong-coupling expansions (\ref{eq: upper boundary}) and (\ref
{eq: lower boundary}) agree with the exact solution (\ref{eq: inf-d delta})
when the latter
is expanded out to third order in $x$, providing an independent check of the
algebra.  Note further, that the exact
solution uniquely determines the expansion coefficients of the powers
of $x$ that do not involve inverse powers of $d$ and the perturbation
expansion is only required to determine the $1/d$ corrections.

The strong-coupling expansion for the $x$, $\mu$
phase diagram in one dimension is compared to the
QMC results of Batrouni et al. \cite{Scalettar}
in Figure 1.
The solid lines indicate the phase boundary between the Mott-insulator phase
and the superfluid phase at zero temperature
as calculated from Eq.~(\ref{eq: upper boundary})
and Eq.~(\ref{eq: lower boundary}).
The solid circles are the results of the QMC calculation\cite{Scalettar} at
a small but finite temperature ($T\approx U/2$). The dotted line
is an extrapolation from the series calculation that will be described below.
Note that the overall agreement of the two
calculations is excellent.
For example, the critical value of the hopping matrix element for the first
Mott lobe ($n_0$) is
$x_{\rm crit} = 0.215$, while the QMC calculations
found\cite{Scalettar} $x_{\rm crit} = 0.215 \pm 0.01$.
A closer examination of Fig. 1 shows a systematic deviation of the lower
branch for larger values of $x$.  We believe that this is most likely a
finite-temperature effect, since the Mott-insulator
phase becomes more stable at higher temperatures \cite{KampfZimanyi}, and the
systematic errors of the QMC
calculation due to finite lattice size and finite Trotter error are
easily controlled\cite{Scalettar2}.

It is known from the scaling theory of Fisher et al. \cite{Fishers} that the
phase transition at the tip of the Mott lobe is in the universality class
of the $(d+1)$ dimensional $XY$ model.
Although a finite-order perturbation theory cannot describe the physics of
the tricritical point correctly, we find that the density fluctuations
dominate the physics of the phase transition even close to the tricritical
point.
Note how the Mott lobes have a cusp-like structure in one dimension, mimicking
the Kosterlitz-Thouless behavior of the critical point.

Figure 2 (a) presents the strong-coupling expansion for the
$x$, $\mu$ phase diagram in two dimensions.
For comparison, the tricritical point of the first
Mott-insulator lobe as obtained by
the QMC simulations of
Krauth and Trivedi \cite{Krauth3} is marked by a solid circle with error
bars (the chemical potential for the tip of the Mott lobe was not reported in
Ref.~\onlinecite{Krauth3}, so we fixed it to be $\mu_{crit}$).   The solid
line is the strong-coupling expansion truncated to
third order, while the dotted line is an extrapolation described below.
Their simulation gives a critical value of $x_{\rm crit} = 0.122\pm 0.006$,
whereas our calculation yields $x_{\rm crit} \approx 0.136$
which is in reasonable agreement. Note that the qualitative shape
of the Mott lobes has changed from one dimension to two dimensions,
mimicking the power-law critical behavior of the $XY$ model
in three or larger dimensions.

Figure 2 (b) shows the corresponding figure for the $n_0\rightarrow\infty$
limit corresponding to the quantum rotor model.  The QMC results are
from van Otterlo and Wagenblast\cite{vanOtterlo}.  The horizontal axis has
been rescaled to $y_{\infty}=\lim_{n_0\rightarrow\infty}n_0x$.  We believe
that the relatively large difference between the QMC and the strong-coupling
perturbation theory arises from the use of the Villain approximation in the
QMC simulations.

Finally the strong-coupling expansion is compared to the exact calculation
in infinite dimensions \cite{Fishers}.
In infinite dimensions, the hopping matrix element must scale inversely with
the dimension \cite{MuellerHartmann}, $t=t^*/d$, $t^* = \text{finite}$,
producing the mean-field-theory result of Eq. (\ref{eq: inf-d delta}).
In Figure 3 the strong-coupling expansion
(solid line) is compared to the exact solution (dashed line) and to an
extrapolated solution (dotted line) which will be described below.
Even in infinite dimensions, the agreement of the strong-coupling expansion
with the exact results is quite good.

As a general rule, the truncated strong-coupling expansions appear to be more
accurate in {\it lower} dimensions, which implies that the density fluctuations
of the bose Hubbard model
are also more important in lower dimensions.

At this point we turn our attention to techniques which enable us to
extrapolate the strong-coupling expansions to infinite order in hopes of
determining a more accurate phase diagram.  The simplest method is called
critical-point extrapolation.  The critical point
$(\mu_{\rm crit},x_{\rm crit})$ is calculated at each order ($m$) of the
strong-coupling expansion and is extrapolated to infinite order $(m\rightarrow
\infty)$. The ansatz that the extrapolation is linear in $1/m$ can be checked
by
determining the correlation coefficient $r$ of the critical points (a value
of $|r|$ that is near 1 indicates a linear correlation).  The
correlations are found to be most linear for large dimensions ($|r|=0.99999$
in infinite dimensions for the first Mott lobe) but remain fairly linear
even in one dimension ($|r|>0.995$ for the $x_{\rm crit}$ extrapolation and
$|r|>0.95$ for the $\mu_{\rm crit}$ extrapolation).  Since the second-
and third-order expansions are expected to be more accurate than the
first-order
calculation, we adopt the following strategy for performing the extrapolations:
the results of the second- and third-order expansions are extrapolated to $m
\rightarrow\infty$ to determine the estimate for the critical point, and the
results of the first, second and third orders are then extrapolated to
$m\rightarrow\infty$ in order to estimate the error in the critical point, and
to test the linear-extrapolation hypothesis.  The error estimate is chosen to
be 1.5 times as large as the difference between the two different
extrapolations.  Figure 4 plots the
critical hopping matrix elements $x_{\rm crit}$ versus $1/m$ for the
infinite-dimensional case and $n_0=1,2,3$.  The solid dots are the results of
the strong-coupling expansion truncated to $m$th order and the solid line
is the linear extrapolant.  The open circles are the exact solutions
from Eq. (\ref{eq: inf-d xcrit}).  Note that although the linear correlation
coefficient is very close to 1, the error in the critical point is
about $2\%$.  The results for the critical-point extrapolation are recorded
in Table I.

The critical-point extrapolation does not yield any information on the shape
of the Mott lobes, but only determines the critical point.  An alternate
extrapolation technique, called the chemical-potential extrapolation method
will determine an extrapolated Mott-phase lobe and critical point.  The
idea is to fix the magnitude of the hopping matrix elements and determine
the value of the chemical potential from Eqs. (\ref{eq: upper boundary}) and
(\ref{eq: lower boundary}) for the upper and lower branch of the Mott lobe.
The chemical potential is determined from a first, second, and third-order
calculation and then extrapolated to infinite order assuming the
ansatz of a linear dependence
upon $1/m$.  This procedure determines an extrapolated Mott lobe that should
be more accurate than the truncated strong-coupling series.  The result for
the infinite-dimensional case is presented as a dotted line in Fig. 3.
Note that the critical point is not determined as accurately by this technique
as it was in the critical-point extrapolation method.  The chemical-potential
extrapolation method fails in one dimension since the extrapolated branches
of the extrapolated Mott lobe do not close.

A third approach is to use the results of the scaling theory\cite{Fishers}.
The critical point is that of a $(d+1)$-dimensional $XY$ model, and therefore,
has a Kosterlitz-Thouless shape in one dimension and a power-law shape
in higher dimensions.  Examination of the exact result for infinite dimensions
(\ref{eq: inf-d delta})
leads one to propose the following ansatz for the Mott lobe in $d\ge 2$
\begin{equation}
{\mu\over U}-n_0=A(x)\pm B(x)(x_{\rm crit}-x)^{z\nu}
\label{eq: mu ansatz}
\end{equation}
where $A(x)\equiv a+bx+cx^2+...$ and $B(x)\equiv \alpha +\beta x+\gamma x^2 +
...$ are regular functions of $x$ (that should be accurately approximated by
their power-series expansions) and $z\nu$ is the critical exponent for the
$(d+1)$-dimensional $XY$ model.  In the unconstrained-scaling-analysis
extrapolation method the exponent $z\nu$ is determined by the strong-coupling
expansion in addition to the parameters $a,b,c$ and $\alpha,\beta,\gamma$.
This provides a perturbative estimate of the exponent $z\nu$ which can be
checked against its well-known values.
In the constrained-scaling-analysis extrapolation method $z\nu$ is
fixed at its predicted values\cite{Fishers} of $z\nu\approx 2/3$ in two
dimensions and $z\nu=0.5$ in higher dimensions.  In direct analogy to
Eq. (\ref{eq: mu ansatz}), we propose the Kosterlitz-Thouless form
\begin{equation}
{\mu\over U}-n_0=A(x)\pm B(x)\exp \left [ -{W\over\sqrt{x_{\rm crit}-x}} \right
]
\label{eq: mu ansatz 1d}
\end{equation}
for the constrained-extrapolation-method in one dimension.

When the unconstrained-scaling-analysis extrapolation method is carried out,
one
finds that there is no solution for the critical exponent in one dimension
(which is consistent with Kosterlitz-Thouless behavior),
that in $d=2$ the exponent satisfies $z\nu\approx 0.58$, in $d=3$ the
exponent is $z\nu\approx 0.54$, and in infinite dimensions $z\nu\approx 0.5$.
There is a slight $n_0$ dependence to the exponents that are calculated
in this method, but that arises from the truncation of the series to such
a low order.  In general, the unconstrained extrapolation method produces
an accuracy of about $15\%$ in the exponent $z\nu$, and the method appears
to work best in higher dimensions.

The results for the constrained-extrapolation method are plotted with a
dotted line in Fig. 1 for the one-dimensional case.  The values of the
critical points are $(\mu_{crit}=0.186, x_{crit}=0.265)$, $(1.319,0.155)$,
and $(2.371,0.111)$
for $n_0=1,2,3$, respectively.  These critical points occur at larger values
of $x$ than predicted by the QMC simulations\cite{Scalettar}, but it is
difficult to gauge whether the extrapolated series expansions are more or
less accurate than the QMC simulations because of the finite-temperature
effects in the latter.
The constrained-extrapolation results in two dimensions are plotted with
a dotted line in Fig. 2 (a) and (b).  The values of the critical points are
$(0.375,0.117)$, $(1.426,0.069)$, $(2.448,0.049)$ for $n_0=1,2,3$,
respectively.  The agreement with the QMC simulations\cite{Krauth3} is
excellent.  Similarly, the extrapolated critical point for the $2-d$ rotor
model is $(0.5,0.171)$ which also agrees well with the QMC.  In this
latter case [Fig.~2~(b)] the errors between the extrapolated series expansion
and the QMC can be traced to the use of the Villain approximation in the
latter.
When the constrained-scaling-analysis extrapolation method is applied to the
infinite-dimensional case the result is indistinguishable from the exact
solution when the two are plotted on the same graph.

The extrapolation techniques work best in {\it higher} dimensions
virtually producing the exact result in infinite-dimensions.  This gives us
confidence that the extrapolated results of the series expansions can
produce numerical answers that are at least as accurate as the QMC simulations.

\section{The disordered case}
The most common type of disorder distribution that has been considered in
relationship to the ``dirty'' boson problem is the Anderson model (continuous
disorder distribution).  In the Anderson model the distribution
$\rho(\epsilon)$ for the on-site energies $\{\epsilon_i\}$
is continuous and flat, satisfying
\begin{equation}
\rho(\epsilon )=\theta(\Delta-\epsilon)\theta(\Delta+\epsilon)
{1\over 2\Delta}\quad .
\label{eq: anderson}
\end{equation}
The symbol $\Delta$ denotes the
maximum absolute value that the site energy $\epsilon_i$ can assume for a
given (bounded) distribution $(|\epsilon_i|\le \Delta U)$.  This
disorder distribution is symmetric $[\rho(-\epsilon)=\rho
(\epsilon)]$ and in particular it satisfies $\sum_i\epsilon_i=0$.
The results presented in this contribution are {\it insensitive} to the actual
shape of the disorder distribution; all we require is a symmetric distribution
with $|\epsilon_i|\le \Delta U$.

We begin by reexamining the exact solution of the infinite-range-hopping
model\cite{Fishers}.  If all energies are measured in units
of the boson-boson repulsion $U$, then the analysis of
Ref.~\onlinecite{Fishers}
derives an equation that relates the hopping matrix elements to the chemical
potential at the Mott phase boundary
\begin{equation}
{1\over 2x} = \int_{-\infty}^{\infty} \left [
-{n_0+1\over y+\epsilon}+{n_0\over y+1+\epsilon} \right ] \rho(\epsilon )
d\epsilon\quad ,
\label{eq: inf-d disorder}
\end{equation}
with $y\equiv -n_0+\mu/U$ the chemical potential and $\rho (\epsilon )$
the disorder distribution.  This solution assumes that the phase transition
from the Mott phase to the bose glass is a {\it second-order} phase transition.

When Eq. (\ref{eq: inf-d disorder}) is solved for the Anderson model
distribution (\ref{eq: anderson}), one finds that the chemical potential
for the lower branch of the Mott lobe
behaves like $y=-\Delta-\exp[-1/2x(n_0+1)]$ for small
$x$.  This result is {\it nonperturbative} in the hopping matrix elements and
cannot be represented by a simple perturbative theory about $x=0$.  The reason
why this happens is that the infinite-range-hopping model has no localized
states for any disorder distribution (however, this statement does depend on
the disorder distribution\cite{zimanyi}).  Localized states can occur in the
infinite-dimensional limit at the tails of the distribution.  Therefore we
expect that the transition will have a different qualitative character on
a hypercubic lattice with nearest-neighbor interactions.
In fact, the perturbative arguments given in the
introduction, show that the phase boundaries have the {\it same} slope
as they did in the pure case.
Furthermore, we expect that the transition to be first order  at the tip of the
Mott lobe because
the states that the bosons initially occupy in the compressible phase are
localized within the rare regions of the lattice (where the
site energies are all equal to $-\Delta U$) implying that there is no diverging
length scale at the transition.
The perturbative expansion for the energy of the Mott phase is unchanged
from Eq. (\ref{eq: EMott}) in the presence of disorder (if the disorder
distribution satisfies $\sum_i\epsilon_i=0)$, while
the defect phases have a trivial dependence upon disorder (in the thermodynamic
limit)---the energy for a particle or hole
defect state is shifted by $-\Delta U$, so the effect of the disorder is
simply to shift the Mott-phase boundaries inward by $\Delta U$.
The critical point, where the Mott phase disappears, is no longer described
by a second-order critical point [in which the slope of $\mu(x)$ becomes
infinite at $x_{crit}$] but rather is described by a first-order critical point
[in which the slope of $\mu(x)$ changes discontinuously at $x_{crit}$].

In the thermodynamic limit one can always find a rare region
of arbitrarily large extent which guarantees the existence of the first-order
transition, but the density of these rare regions
is an exponentially small function of their size.  For this reason the
compressibility at the Mott-phase boundary will also be exponentially small.
Finite-size effects play a much more important role in the disordered case:
{\it it is impossible to determine the Mott-phase boundary accurately
in the thermodynamic limit by scaling calculations performed on small
lattices, because the lattice size must be large enough to contain
rare regions within which the bosons can be delocalized.} (Finite-size effects
can be studied with the strong-coupling expansion which is given to third order
in the single-particle matrix $S$ in the appendix.)

The most accurate way of calculating the Mott phase boundary is then to take
the
results of the constrained-scaling-analysis extrapolation for the pure case
and shift the branches by the strength of the disorder.  This is plotted
in Figure 5 for the one-dimensional case and two different values of
disorder ($\Delta=0,0.25$). The thermodynamic limit is represented
by the solid line for the pure case, and dotted lines for the disordered
case, while the dashed line is the result of an Anderson-model
disorder distribution on a finite lattice with 256 sites.  The QMC results of
Batrouni, {\it et.~al.}\cite{Scalettar} correspond
to lattice sizes ranging from 16 sites to 256 sites (the disorder parameter
is $\Delta=0$ for the solid dots and $\Delta=0.25$ for the open dots).
The Mott phase is stabilized on finite-sized systems because they do not
possess the rare regions needed to correctly determine the Mott phase
boundary.  This is clearly seen in the QMC results, which predict a much
larger region for the Mott phase than the strong-coupling perturbation theory
does in the thermodynamic limit.  The perturbative results for a finite system
are much closer to the QMC results as expected. (Note that the finite-size
calculations have not been extrapolated, so they should underestimate the
stability of the Mott phase in one dimension which is clearly seen in
Figure~5.)
Also the slope of the phase boundary approaches zero (as $x\rightarrow 0)$
for the finite-size systems\cite{Scalettar}.

Figure 6 plots the Mott-phase diagram for the disordered bose Hubbard model
in two dimensions and one value of the disorder ($\Delta=0,0.182$)
in the thermodynamic limit.  The solid dot (with error bars)
is the QMC result\cite{Krauth3} (for the pure case with $\Delta=0$) and the
open dot is the disordered case $(\Delta=0.182$).  Note that in dimensions
larger than one, the finite-size effects for the tip of the Mott lobe are
not as strong as they are in one dimension.

We compare in Figure 7 the differences between the infinite-dimensional lattice
and the infinite-range-hopping model of Ref.~\onlinecite{Fishers}.
The solid lines correspond to the exact solution
with no disorder, the dotted lines are the infinite-dimensional lattice
strong-coupling expansion with disorder
($\Delta=0.2$),
and the dashed lines are the exact solution of the infinite-range-hopping
model.
In the case of disorder, the first-order nature of the transition is evidenced
by the jump in the slope of the Mott phase boundary at the tip of the lobe.
The second-order phase
boundaries predict a more stable Mott phase, and their slopes all approach
zero as $x\rightarrow 0$.  We expect in the region in between the (first-order)
infinite-dimensional phase
boundary and that of the (second-order) infinite-range-hopping model that the
compressibility will be exponentially small, and will only become
sizable as the second-order phase boundary is crossed.

Because the Mott-phase to bose-glass phase transition is first order
for the disordered case, and since the bose-glass to superfluid transition
is always second order (because it involves a collective excitation that
extends through the entire lattice), it is quite unlikely that there would
ever be a region where the Mott phase has a transition directly to the
superfluid.  {\it The presence of the Lifshitz rare regions strongly supports
the picture that the Mott phase is entirely enclosed within the bose-glass
phase.}  This result is {\it independent} of any perturbative arguments, since
the rare regions must dominate the Mott to bose-glass transition in the exact
solution too.

Finally, we calculate the dependence of the critical value of $x$ at the
tip of the Mott-phase lobe, as a function of the disorder strength $\Delta$.
Figure 8 plots this value of $x_{crit}(\Delta)/x_{crit}(0)$ versus $\Delta$
for the one-, two-, and infinite-dimensional cases.  The plot is limited to
the lowest Mott-phase lobe with $n_0=1$.  Since the one-dimensional Mott
phase lobes have a cusp-like behavior that is removed when disorder is added
to the system, we expect $x_{crit}$ to decrease very rapidly for small
disorder.  This effect is sharply
reduced in higher dimensions.  In the strong-disorder
limit, the phase diagram is dominated by
the first-order terms in the perturbative expansion, which have a trivial
dependence on the dimensionality, but the slopes curves of the curves are
unequal because $x_{\rm crit}(0)$ depends strongly upon the dimensionality.

\section{Conclusion}

We have developed a strong-coupling perturbation-theory approximation to the
bose Hubbard model on a bipartite lattice.  The perturbative results can
be extrapolated in a number of different ways which either do or do not take
into account the scaling theory of the critical point at the Mott tip.
We find that a perturbative expansion thru third order rivals the accuracy
of the QMC simulations, and is likely the best method for determining the
Mott phase boundary of these interacting bose systems.

We treated two different cases: the pure case and the disordered case.  In
the pure case the tip of the Mott lobe satisfies a scaling relation because
it corresponds to a second-order phase transition in a $d+1$-dimensional
XY model.  This is because the compressible phase is also superfluid which
implies there is a diverging length scale at the phase transition.
Calculations in the pure case are insensitive to finite-size effects.
In the disordered case we argued that the tip of the Mott phase lobe
corresopnds to a first-order phase transition because the initial
single-particle excitations are localized into the rare regions of the
Lifshitz tails for any bounded disorder distribution.  As a result there is
a kink in the Mott phase boundary since the slope of $\mu(x)$ has a
discontinuous jump at the tip of the lobe.  In this case, there are strong
finite-size effects because ``typical'' disorder distributions on finite
lattices do not have Lifshitz tails.

The results of these perturbative calculations have been compared to
the available QMC simulations.  We find a remarkable agreement between the
two and are unable to determine which method is more accurate in
a quantitative determination of the phase boundaries.

The perturbation theory described here falls short in one aspect---it is unable
to determine the bose-glass--superfluid phase transition in the disordered
case.  It is possible that such a calculation could be performed, but since
the particle density at which it occurs is a priori not known, and since
the superfluid susceptibility diverges in the bose-glass phase, such a
calculation may be problematic.

\acknowledgments
We would like to thank M. Fisher, Th. Giamarchi,
M. Jarrell, M. Ma, A van Otterlo,
R. Scalettar, R. Singh, and G. Zimanyi
for many useful discussions.  We would especially like to thank M. Ma for
pointing out that the Mott to bose-glass phase transition is first order
in the disordered case.  Initial stages of this work were carried out by
J. K. F. at the University of California, Davis in 1994 and were completed
during a visit to
l'\'Ecole Polytechnique F\'ed\'erale de Lausanne in June 1995.
JKF would like to thank the Office of Naval Research (under Grant No.
N00014-93-1-0495) for support while at UC Davis, and would like to thank the
Donors of the Petroleum Research Fund, administered by the American Chemical
Society (ACS-PRF-29623-GB6) for support while at Georgetown.

\appendix
\section{Finite-size effects for the disordered case}

The Rayleigh-Schr\" odinger perturbation theory for the case with disorder is
straightforward, but rather tedious.  In the thermodynamic limit, the
perturbative expansion simplifies because it is dominated by the rare regions
of the lattice.  For a finite-sized system, the perturbative results are more
complicated.  We summarize here the main results for a third-order
strong-coupling expansion in the presence of disorder.

The only assumptions that are made about the disorder distribution is that it
is bounded and symmetric, so that $|\epsilon_i|\le \Delta U$ and
$\sum_i\epsilon_i =0$.  Restriction is also made to bipartite lattices.

The upper phase boundary for the Mott to bose-glass transition is found by
solving the equation
\begin{eqnarray}
\delta^{(part)}(n_0)&={\Lambda\over U}(n_0+1)+{1\over 2}{Zt^2\over U^2}
n_0(5n_0+4)-\sum_{ijk}f_i{t_{ij}t_{jk}\over U^2}f_kn_0(n_0+1)\cr
&-n_0(n_0+1)\Biggr [ \sum_{ijkl}f_i{t_{ij}t_{jk}t_{kl}\over U^3}f_l(3n_0+2)
+{\Lambda\over U}\sum_{ijk}f_i{t_{ij}t_{jk}\over U^2}f_k(n_0+1)\cr
&-{Zt^2\over U^2}\sum_{ij}f_i{t_{ij}\over U}f_j(9n_0+6)+{t^2\over U^2}
\sum_{ij}f_i{t_{ij}\over U}f_j(4n_0+2)-{\Lambda Zt^2\over U^3}({11\over 4}n_0
+{5\over 2})\cr
&-\sum_{ijk}{(\epsilon_i-\epsilon_j+\epsilon_k)\over U}
f_i{t_{ij}t_{jk}\over U^2}f_k \Biggr ]\cr
&+n_0\left [ \sum_{ij}{\epsilon_i\over U}{t_{ij}^2\over U^2}f_j^2({3\over 2}
n_0+2)
-{Zt^2\over U^2}\sum_i{\epsilon_i\over U}f_i^2({17\over 4}n_0+{9\over 2})
\right ]\quad ,
\label{eq: appendix1}
\end{eqnarray}
where $f_i$ is the minimal eigenvector of the single-particle matrix
$S_{ij}^{(part)}=
-t_{ij}+\epsilon_i\delta_{ij}/(n_0+1)$ and $\Lambda$ is its corresponding
eigenvalue.  The identity $\sum_{ij}t_{ij}^2f_j^2=Zt^2$
was needed in deriving the above result.  A similar calculation yields
\begin{eqnarray}
\delta^{(hole)}(n_0)&=-{\tilde\Lambda\over U}n_0-{1\over 2}{Zt^2\over U^2}
(n_0+1)(5n_0+1)+\sum_{ijk}g_i{t_{ij}t_{jk}\over U^2}g_kn_0(n_0+1)\cr
&-n_0(n_0+1)\Biggr [ - \sum_{ijkl}g_i{t_{ij}t_{jk}t_{kl}\over U^3}g_l(3n_0+1)
-{\tilde\Lambda\over U}\sum_{ijk}g_i{t_{ij}t_{jk}\over U^2}g_kn_0\cr
&+{Zt^2\over U^2}\sum_{ij}g_i{t_{ij}\over U}g_j(9n_0+3)-{t^2\over U^2}
\sum_{ij}g_i{t_{ij}\over U}g_j(4n_0+2)+{\tilde\Lambda Zt^2\over U^3}
({11\over 4}n_0+{1\over 4})\cr
&-\sum_{ijk}{(\epsilon_i-\epsilon_j+\epsilon_k)\over U}
g_i{t_{ij}t_{jk}\over U^2}g_k \Biggr ]\cr
&+(n_0+1)\left [ \sum_{ij}{\epsilon_i\over U}{t_{ij}^2\over U^2}g_j^2({3\over
4}
n_0+{1\over 4})
-{Zt^2\over U^2}\sum_i{\epsilon_i\over U}g_i^2({7\over 2}n_0+{1\over 2})
\right ]\quad ,
\label{eq: appendix2}
\end{eqnarray}
for the lower phase boundary of the Mott to bose-glass transition.  Here we
have that $g_i$ is the minimal eigenvector of the single-particle matrix
$S_{ij}^{(hole)}=
-t_{ij}-\epsilon_i\delta_{ij}/n_0$ and $\tilde\Lambda$ is its corresponding
eigenvalue.

In the thermodynamic limit we know that the minimal eigenvalue occurs in the
rare regions where the disorder is constant and equal to its extreme value.
The ground-state eigenvector is delocalized within the rare region (to minimize
it's kinetic energy) and localized to the rare region (to minimize its disorder
energy).  Such an eigenvector is now {\it separately} an eigenvector of the
kinetic-energy matrix and of the disorder matrix, so we have $-\sum_{ij}
t_{ij}f_j=\lambda f_i$, $\sum_j\epsilon_i\delta_{ij}f_j=-\Delta U f_i$,
and $\Lambda=\lambda-\Delta U/(n_0+1)$ with similar formulae for the
$g$ eigenvector.  Plugging these thermodynamic limits into
Eqs.~(\ref{eq: appendix1}) and (\ref{eq: appendix2}) then yields the result
that the Mott phase boundaries are only shifted uniformly by $\Delta U$, and
all higher-order dependence on the disorder vanishes.

\mediumtext
\begin{table}
\caption{
   Results for the critical-point extrapolation method described in the text.
   The critical point is recorded for the first three Mott lobes in one, two,
   three and infinite dimensions.  Where possible the results from other
   calculation techniques are displayed in the last column.}
\label{table1}
\begin{tabular}{ccccc}
dimension & $n_0$ & $\mu_{\rm crit}$ & $x_{\rm crit}$ & $x_{\rm crit}$(exact)\\
\tableline
1 & 1 & 0.255$\pm$0.11 & 0.245$\pm$0.012 & 0.215$\pm$0.010\tablenote{Ref. 2
(Monte Carlo simulation at finite temperature)}\\
  & 2 & 1.359$\pm$0.06 & 0.145$\pm$0.009 & 0.130$\pm$0.020$^{\rm a}$\\
  & 3 & 2.400$\pm$0.04 & 0.103$\pm$0.006 & 0.104$\pm$0.020$^{\rm a}$\\
2 & 1 & 0.388$\pm$0.05 & 0.114$\pm$0.013 & 0.122$\pm$0.006\tablenote{Ref. 7
(Monte Carlo simulation at finite temperature)}\\
  & 2 & 1.435$\pm$0.03 & 0.067$\pm$0.006 &\\
  & 3 & 2.454$\pm$0.02 & 0.048$\pm$0.004 &\\
3 & 1 & 0.400$\pm$0.03 & 0.101$\pm$0.008 &\\
  & 2 & 1.441$\pm$0.02 & 0.060$\pm$0.004 &\\
  & 3 & 2.458$\pm$0.01 & 0.042$\pm$0.002 &\\
$\infty$ & 1 & 0.416$\pm$0.001 & 0.0843$\pm$0.001 & 0.0858\tablenote{Ref. 1
[Exact solution from Eq. (14)]}\\
         & 2 & 1.451$\pm$0.002 & 0.0494$\pm$0.002 & 0.0505$^{\rm c}$\\
         & 3 & 2.465$\pm$0.001 & 0.0351$\pm$0.001 & 0.0359$^{\rm c}$\\
\end{tabular}
\end{table}

\begin{figure}[t]
  \caption{
  The $x$, $\mu$ phase diagram of the bose Hubbard model in one
  dimension ($d = 1$). The solid lines give the phase boundaries of the
  Mott insulator to the superfluid state as determined from
  a third-order strong-coupling calculation. The dotted line is the constrained
  fit to a Kosterlitz-Thouless form.    The circles are the result of the QMC
  calculation of Batrouni {\it et. al.}$^2$.
  }
  \label{fig:1}
\end{figure}

\begin{figure}[t]
  \caption{
  (a) The $x$, $\mu$ phase diagram of the bose Hubbard model in two
  dimensions ($d = 2$). The solid lines give the phase boundaries of the
  Mott insulator to the superfluid state as determined from
  a third-order strong-coupling calculation. The dotted line is the constrained
  fit to a power law form with exponent $z\nu =2/3$.
  The point (with error bars) indicates the tricritical point as
  determined by the QMC calculation of Krauth and
  Trivedi$^7$ (no value for $\mu_{crit}$ was given in Ref.~7).
  (b) The $y_{\infty}$, $\mu$ phase diagram for the quantum rotor model in
  two dimensions.  The solid lines are the perturbative results to third
  order and the dotted lines are the constrained extrapolation fit.  The dots
  are the QMC results of van Otterlo and Wagenblast$^8$.  The disagreement
  between the QMC and the extrapolated results most likely arise from the
  use of the Villain approximation in the former.
  }
  \label{fig:2}
\end{figure}
\begin{figure}[t]
  \caption{
  The $x$, $\mu$ phase diagram of the bose Hubbard model in infinite
  dimensions ($d \rightarrow \infty$).
  The solid lines give the phase boundaries of the
  Mott insulator to the superfluid state as determined from
  a third-order strong-coupling calculation.
  The dashed lines are the result of the mean-field
  calculation of Fisher {\it et. al.}$^1$.
  The dotted lines are the chemical-potential
   extrapolation described in the text.
  }
  \label{fig:3}
\end{figure}

\begin{figure}[t]
  \caption{
   The critical-point extrapolation method in infinite dimensions.  The solid
   circles are the results for $x_{\rm crit}$ calculated at first, second, and
   third order.  The solid line is the linear extrapolation to infinite order.
   The open circles are the exact result.  Three cases are shown $n_0=1$ (top),
   $n_0=2$ (middle), and $n_0=3$ (bottom).  Note that although the three points
   have a correlation coefficient larger than 0.9999 the accuracy of the
   extrapolated critical point is only on the order of $2\%$.
   }
   \label{fig:4}
\end{figure}

\begin{figure}[t]
  \caption{
   Phase diagram in one dimension with disorder.  The perturbative
   approximations are plotted with solid (dotted) lines for the pure
   (disordered) cases in the thermodynamic limit.  Three cases of disorder
   are included: ($\Delta=0.125, 0.25, 0.375$).  The dashed line is the
   perturbative results for a finite system with 256 lattice sites.
   The solid dots are the quantum Monte Carlo results from
   Ref.~2
   in the pure case, and the open dots are for the disordered case
   with $\Delta=0.25$ (the disordered calculations were performed on lattice
   sizes ranging from 16 to 256 sites).
   }
   \label{fig:5}
\end{figure}

\begin{figure}[t]
  \caption{
   Phase diagram in two dimensions with disorder.  The perturbative
   approximations are plotted with solid (dotted) lines for the pure
   (disordered) cases in the thermodynamic limit.  The disorder was set
   equal to $\Delta=0.182$.  The solid (open) dots are the quantum Monte
   Carlo results of Ref.~7 for the pure (disordered)
   cases.  Surprisingly, in two and higher dimensions, the finite-size effects
   in the disordered regime appear to be weaker.
   }
   \label{fig:6}
\end{figure}

\begin{figure}[t]
  \caption{
   Phase diagram in infinite dimensions with disorder.  The perturbative
   approximations are plotted with solid (dotted) lines for the pure
   (disordered) cases in the thermodynamic limit.  The disorder was set
   equal to $\Delta=0.2$.  The dashed line is the exact solution of the
   infinite-range-hopping model from
   Ref.~1.  Note how the Mott phase is more stable
   with the infinite-range calculation, and how it
   has vanishing slope as $x\rightarrow 0$.  Interestingly, the location
   of the tip of the Mott lobe is close for both the infinite-dimensional and
   the infinite-range calculations.
   }
   \label{fig:7}
\end{figure}

\begin{figure}[t]
  \caption{
   Plot of $x_{\rm crit}(\Delta)/x_{\rm crit}(0)$ as a function of dimension.
   The solid line is for one dimension, the dotted line for two dimensions,
   and the dashed line for infinite dimensions.  Note that the intitial
   decrease of $x_{crit}$ is very rapid in one dimension, because of the
   cusp-like shape of the Mott lobe in the pure case, but is much slower
   in higher dimensions (because the tip has a power law behavior in the
   pure case).
   }
   \label{fig:8}
\end{figure}

\newpage

\begin{figure}[t]
\epsfxsize=5.0in
\epsffile{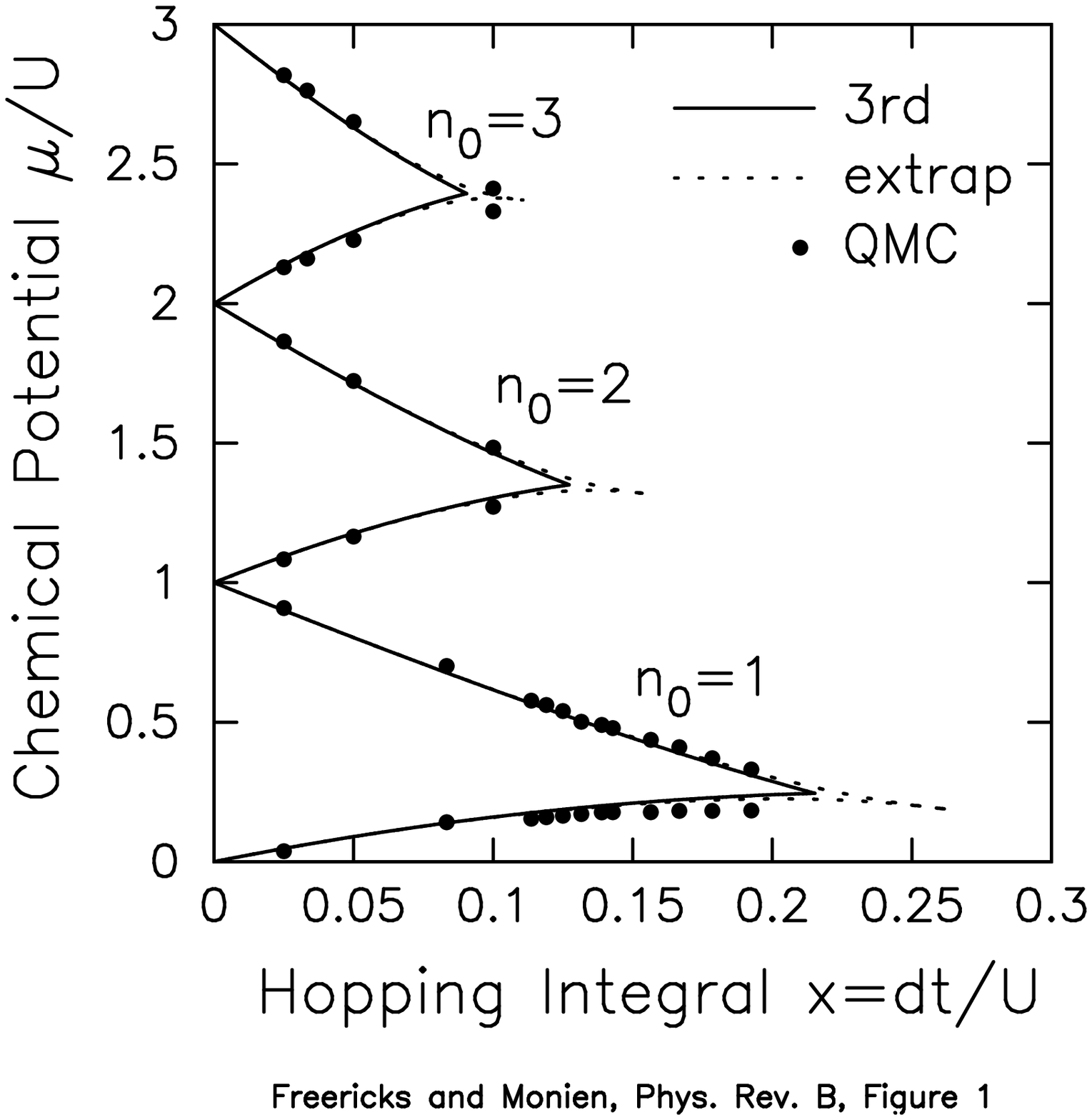}
\end{figure}

\begin{figure}[t]
\epsfxsize=5.0in
\epsffile{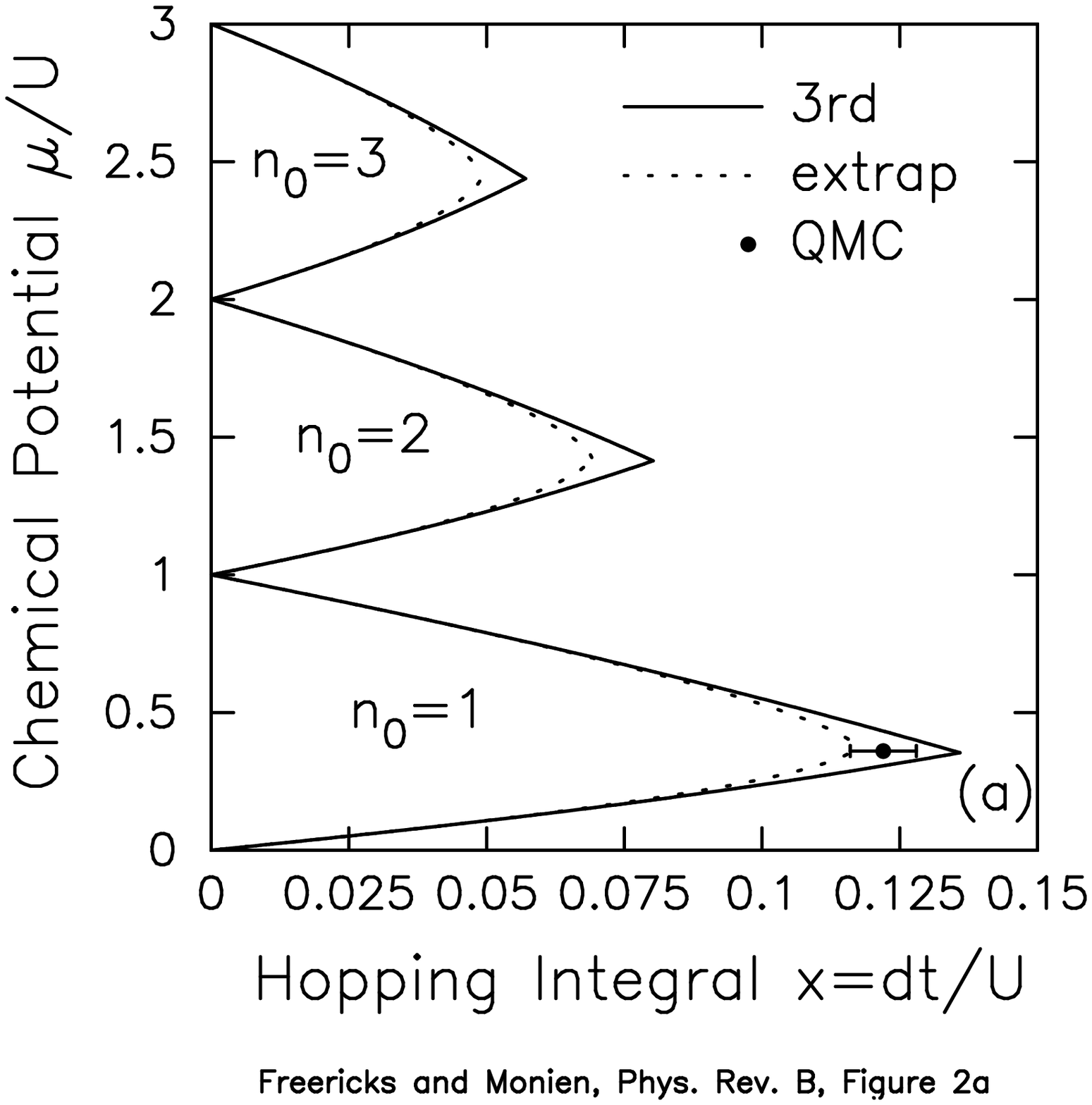}
\end{figure}

\begin{figure}[t]
\epsfxsize=5.0in
\epsffile{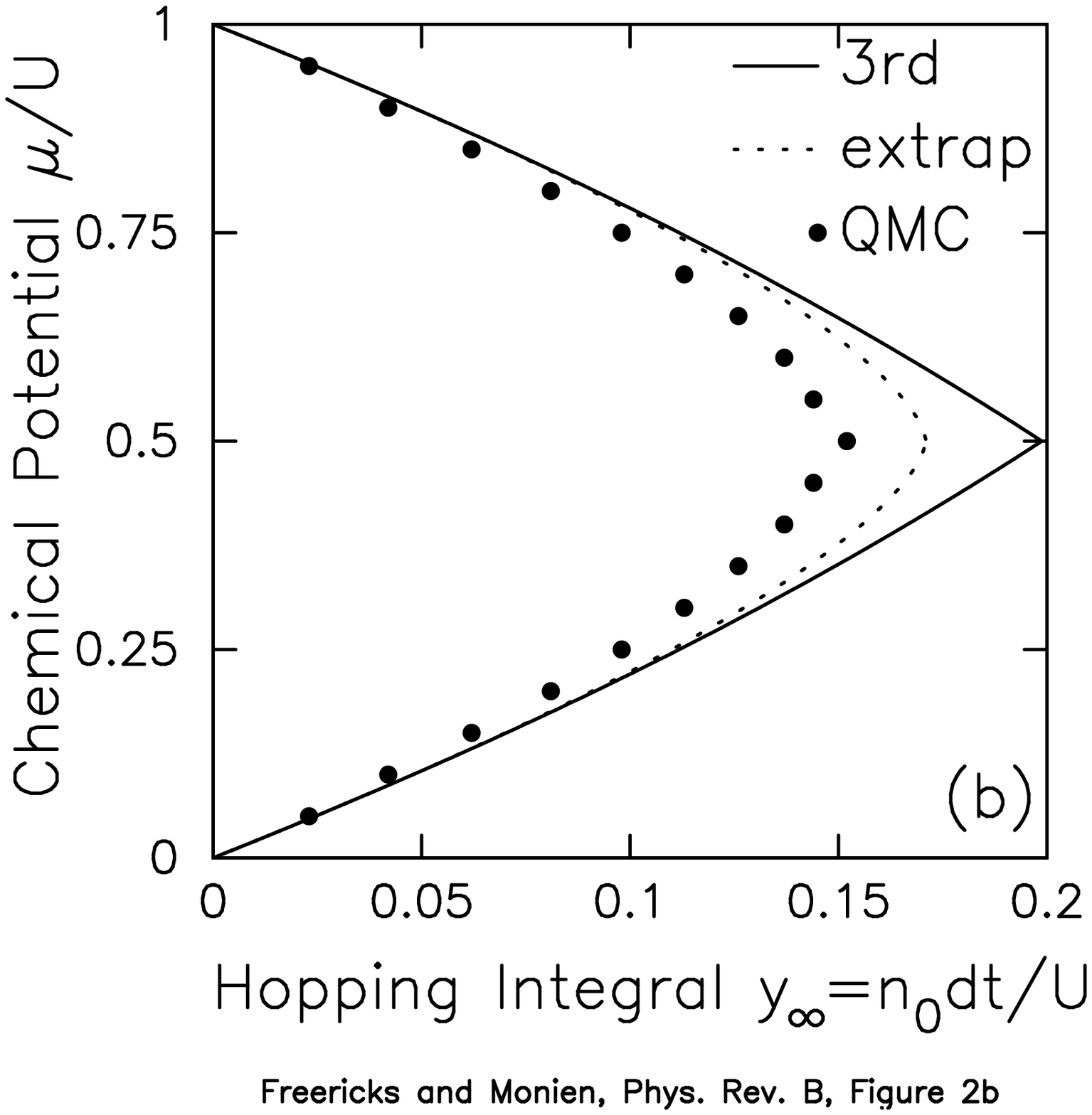}
\end{figure}

\begin{figure}[t]
\epsfxsize=5.0in
\epsffile{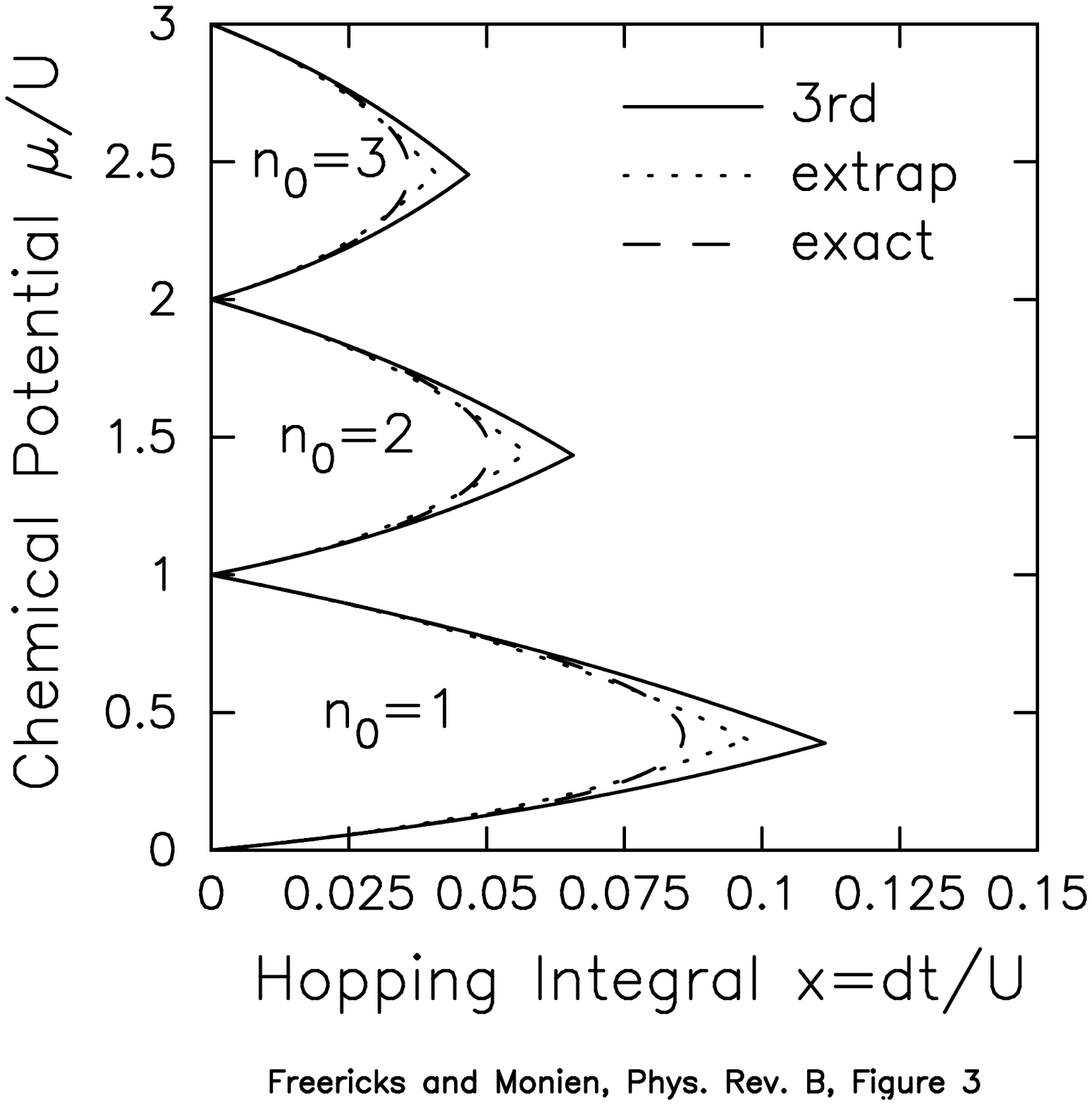}
\end{figure}

\begin{figure}[t]
\epsfxsize=5.0in
\epsffile{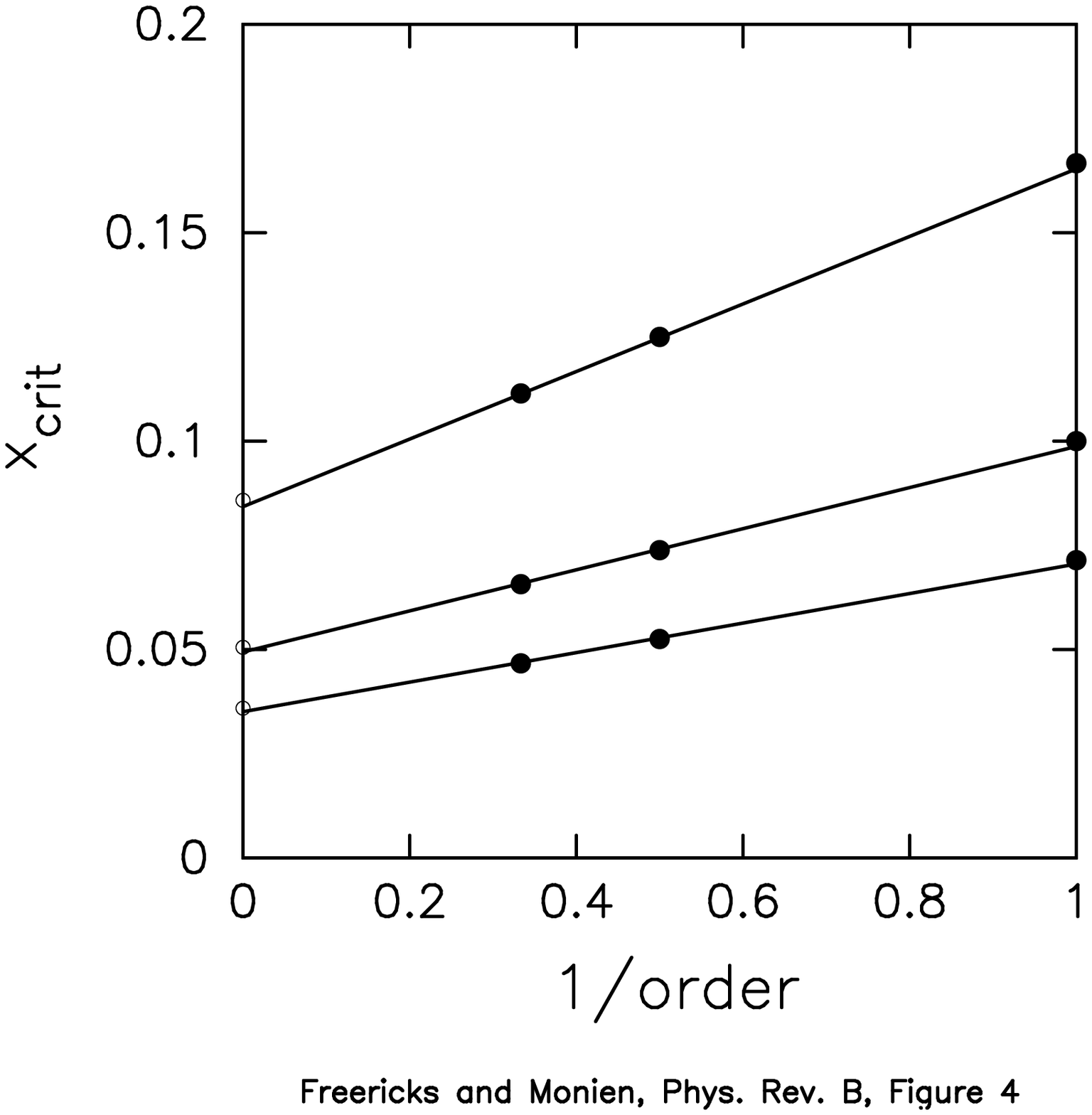}
\end{figure}

\begin{figure}[t]
\epsfxsize=5.0in
\epsffile{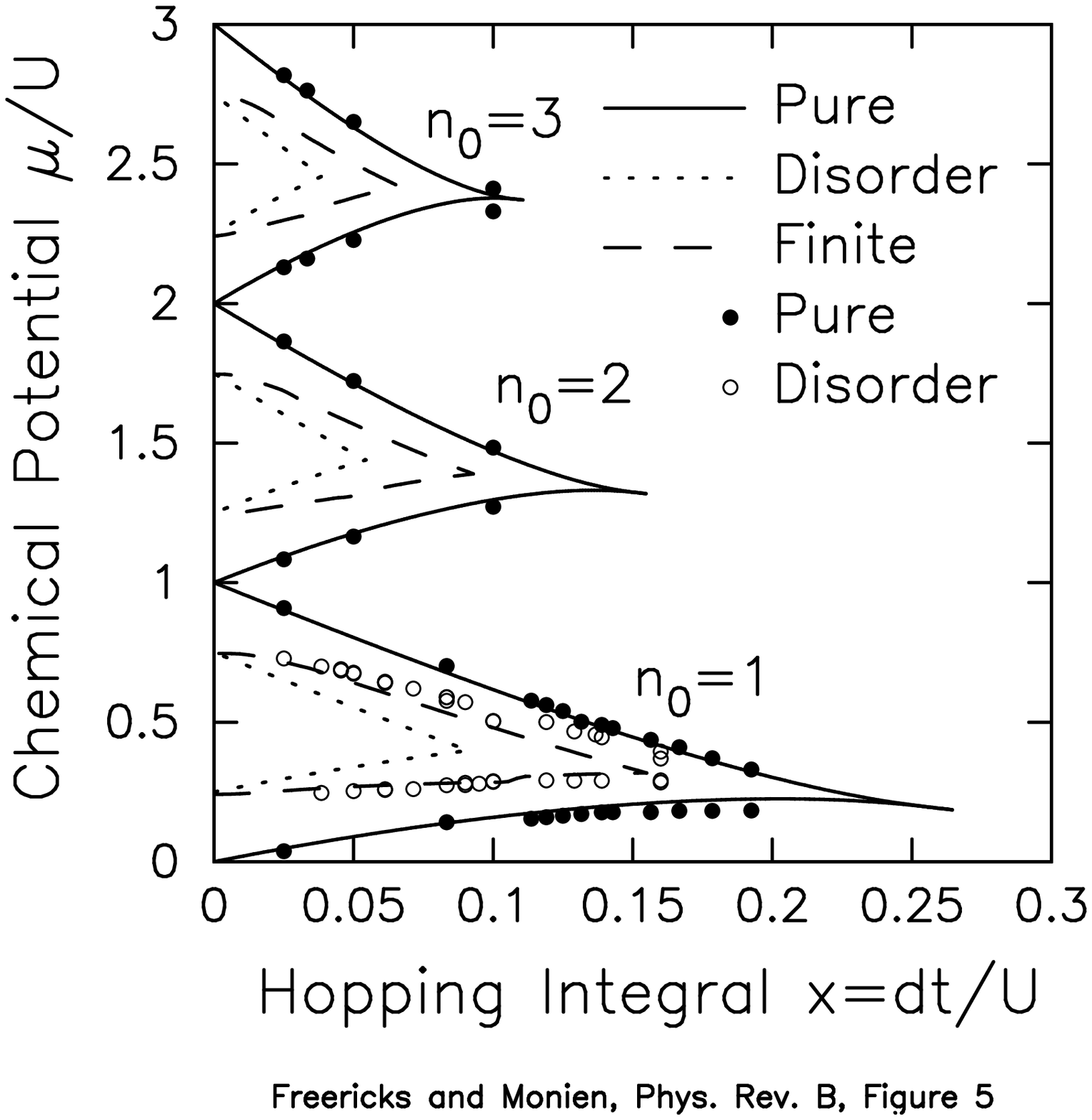}
\end{figure}

\begin{figure}[t]
\epsfxsize=5.0in
\epsffile{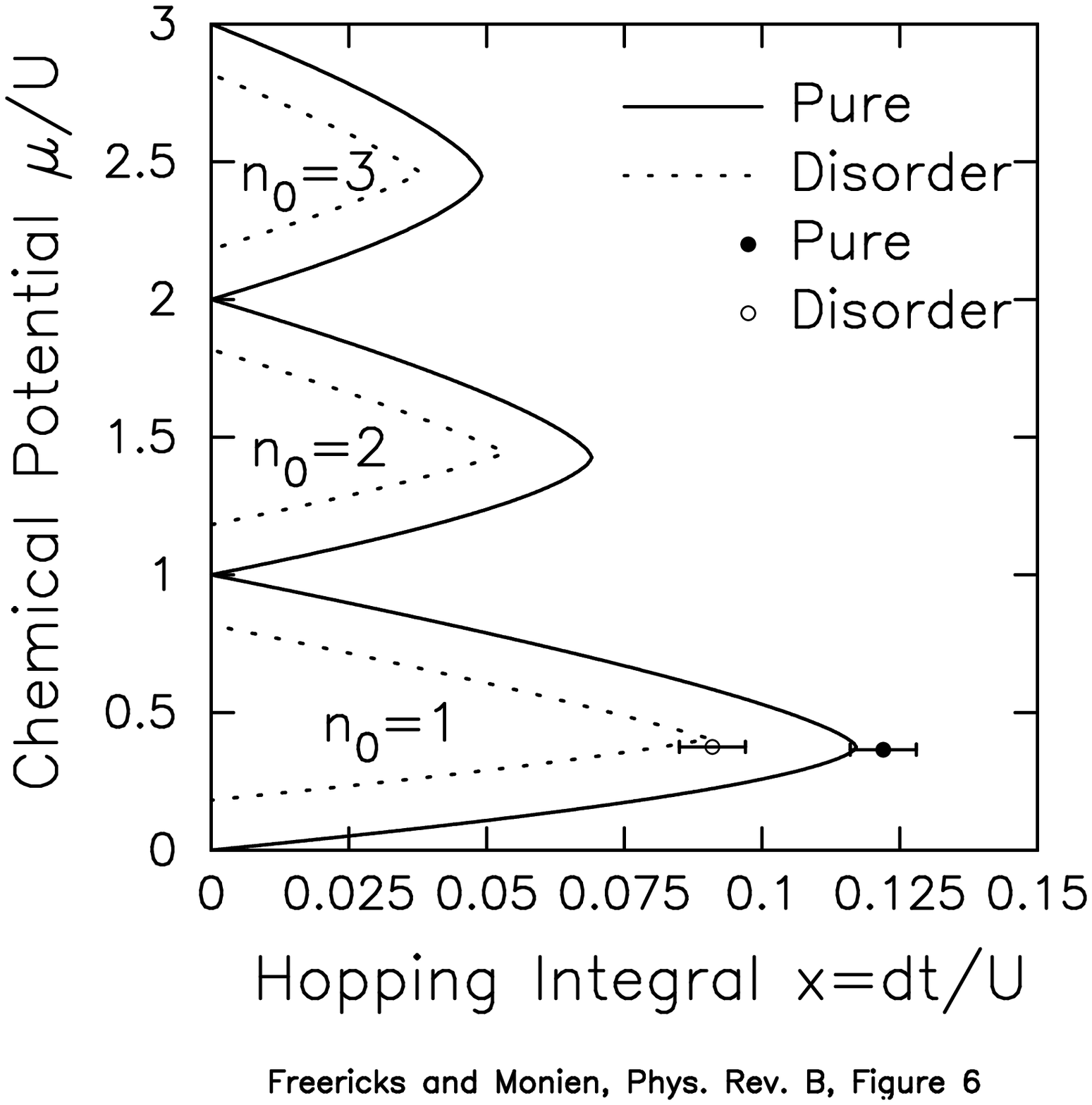}
\end{figure}

\begin{figure}[t]
\epsfxsize=5.0in
\epsffile{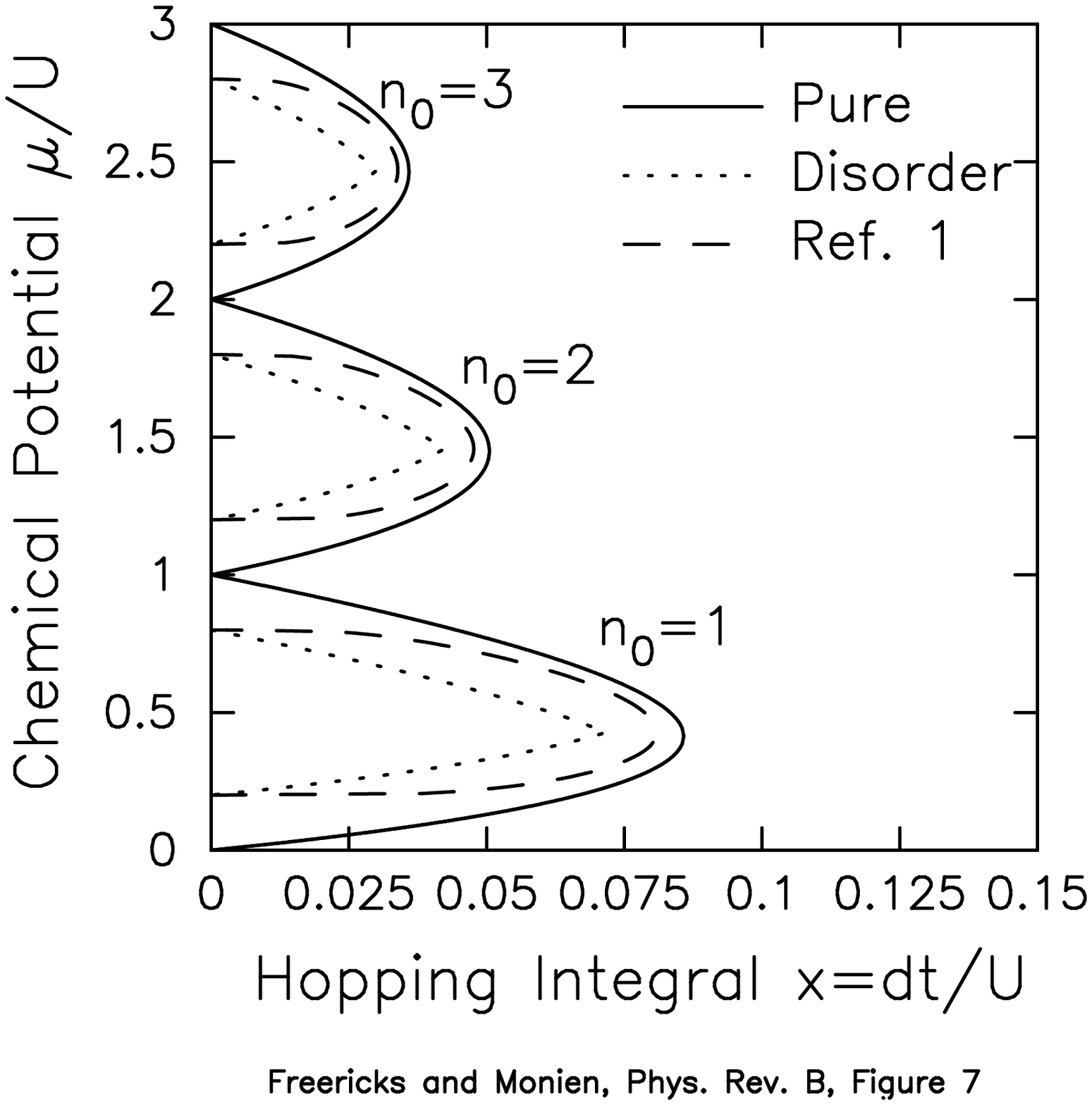}
\end{figure}

\begin{figure}[t]
\epsfxsize=5.0in
\epsffile{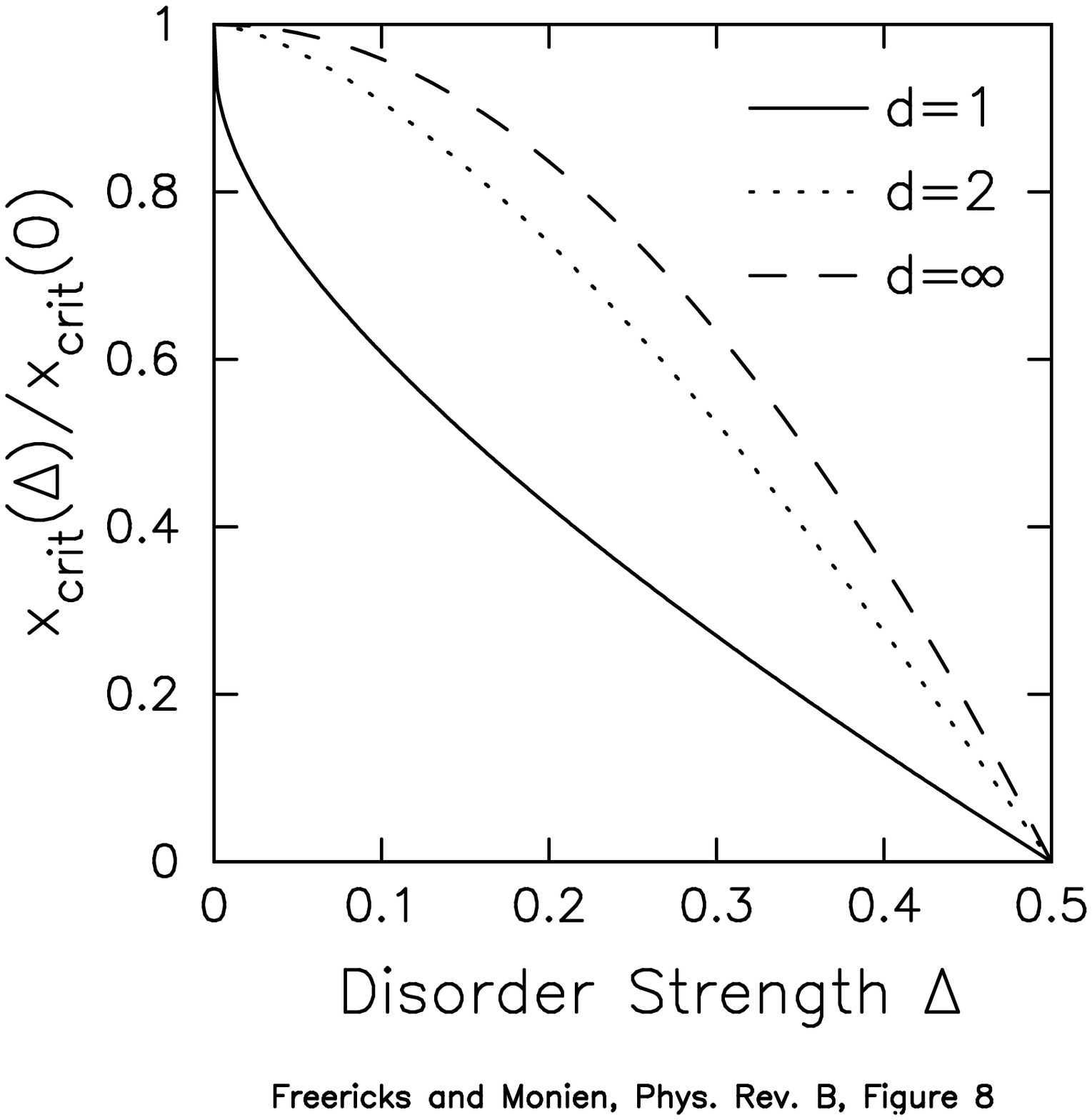}
\end{figure}

\end{document}